\documentclass[10pt,journal,compsoc]{IEEEtran}
\usepackage{CJK}
\usepackage{amssymb}
\usepackage{amsmath}
\usepackage{graphicx, subfigure}
\usepackage{url, cite}
\usepackage{verbatim}
\usepackage{booktabs}
\usepackage{multirow}
\usepackage{bigstrut}
\usepackage{booktabs}

\usepackage{amsmath,cite,graphicx,times,subfigure,url,verbatim}
\usepackage{algorithmic}
\usepackage{algorithm}

\newtheorem{subsec:coding}{subsec:coding}

\usepackage{color}

\renewcommand{\IEEEQED}{\IEEEQEDopen}

\begin{document}

\title{A Sustainable Multi-modal Multi-layer Emotion-aware Service at the Edge}

\author{Long~Hu, Wei~Li, Jun~Yang, Giancarlo~Fortino,~\IEEEmembership{Senior Member,~IEEE,} and Min~Chen,~\IEEEmembership{Senior Member,~IEEE}


\IEEEcompsocitemizethanks{\IEEEcompsocthanksitem Long~Hu, Wei~Li, Jun~Yang, and Min~Chen are with School of Computer Science and Technology, Huazhong University of Science and Technology, Wuhan 430074, China (longhu.cs@gmail.com, weili\_epic@hust.edu.cn, junyang\_cs@hust.edu.cn, minchen2012@hust.edu.cn).
\IEEEcompsocthanksitem Giancarlo~Fortino is with University of Calabria (g.fortino@unical.it).
\IEEEcompsocthanksitem Min~Chen is the corresponding author.}
}

\IEEEtitleabstractindextext{
\begin{abstract}
  Limited by the computational capabilities and battery energy of terminal devices and network bandwidth, emotion recognition tasks fail to achieve good interactive experience for users. The intolerable latency for users also seriously restricts the popularization of emotion recognition applications in the edge environments such as fatigue detection in auto-driving. The development of edge computing provides a more sustainable solution for this problem.
  Based on edge computing, this article proposes a multi-modal multi-layer emotion-aware service (MULTI-EASE) architecture that considers user's facial expression and voice as a multi-modal data source of emotion recognition, and employs the intelligent terminal, edge server and cloud as multi-layer execution environment.
  By analyzing the average delay of each task and the average energy consumption at the mobile device, we formulate a delay-constrained energy minimization problem and perform a task scheduling policy between multiple layers to reduce the end-to-end delay and energy consumption by using an edge-based approach, further to improve the users' emotion interactive experience and achieve energy saving in edge computing.
  Finally, a prototype system is also implemented to validate the architecture of MULTI-EASE, the experimental results show that MULTI-EASE is a sustainable and efficient platform for emotion analysis applications, and also provide a valuable reference for dynamic task scheduling under MULTI-EASE architecture.
\end{abstract}

\begin{IEEEkeywords}
computing offloading, edge computing, emotion recognition, energy saving, multi-modal data
\end{IEEEkeywords}
}

\maketitle

\IEEEpeerreviewmaketitle

\section{Introduction}   \label{sec:intro}
Emotion recognition is a complex process, that is aimed at the individual's emotional state, which means that the emotions corresponding to each individual's behavior are different~\cite{1}.
With the rapid development of artificial intelligence, many tasked related to computer vision, speech recognition, and natural language processing have achieved great success~\cite{image,voice,nlp}, but emotional modeling for individuals still faces great challenges. Presently, the main methods for emotion recognition include speech emotion recognition~\cite{speech}, text emotion recognition~\cite{text}, facial emotion recognition~\cite{face}, gesture emotion recognition~\cite{gest}, etc. As the emotional data source is very rich, the single-modal model can not judge the user's emotion well, so we need multi-modal data to learn the emotion model. In this paper, the emotion analysis is based on multi-modal data such as user's facial expression and voice.

In addition, the emotion recognition task cannot achieve a good interactive experience due to the undesirable interactive latency.
To solve the problem, we introduce edge computing and put forward a multi-layer model for algorithm execution, that is emotional services can be provided at device terminals, network edges and clouds~\cite{yin17t,Fortino19iot}.
The emergent paradigm of edge computing~\cite{edge2,dynamicEdge,f18e} advocates that computational and storage resources can be extended to the edge of the network so that the impact of data transmission latency over the Internet can be effectively reduced for time-constrained applications, i.e., emotion analysis~\cite{ems}.
For example, in auto driving scene, the virtual emotional robot needs to collect the driver's voice, facial expression and other multi-modal emotional data, as well as some surrounding environmental information, including the driver's current road condition information, geographical location, time information, etc., which has strong demands on computational and storage resources, as well as low latency.
Therefore, we deploy some AI algorithms at the edge computing nodes to handle emotion recognition tasks~\cite{learningEdge}. These algorithms usually have lower performance than algorithms deployed in the cloud, but edge computing nodes can quickly feed back results to users.


However, with the widespread deployment of edge computing devices, the energy demand of these devices has increased and started to become a noticeable issues for sustainable development of time-constrained IoT applications~\cite{edge3,yin17s,hu2019fgcs,fiot}. Energy efficiency optimization for such applications becomes more challenging when considering the rapid constant grow of edge devices/sensors~\cite{hao18e,hu2018access}. For example, the current number of IoT devices will rapidly increase from 15 billion to 50 billion by 2020 (according to CISCO), while the number of sensors will increase to as high as 1 trillion by 2030 (according to HP Labs).
In emotion analysis applications~\cite{f19h}, we need multi-modal data for the emotion recognition, such application consumes lots of sensors for acquiring emotion-related data such as facial expression, voice and other physical data. The sustainability of such systems becomes a necessity.
Although local computing consumes more energy, it can significantly minimize the execution latency without additional communication or waiting delay.
Thus, it is critical to make efficient offloading decision between energy consumption of smart mobile devices and execution latency of the corresponding tasks~\cite{yin16g,minOffloading,offloading}.


Above all, a flexible task scheduling strategy must be adopted considering the delay and energy demands\cite{peng2018access,hu2018iotj}. The delay, energy consumption and quality of service (such as the accuracy of emotion recognition in this paper) are different at different layer levels, and the user's sensitivity to the delay of emotion recognition is also different. At the same time, considering the energy consumption problem, we hope to find a kind of compromise scheduling scheme to enable energy consumption to be reduced while satisfying user delay requirements and service quality requirements.
Therefore, in this paper, we propose a multi-modal multi-layer emotion-aware services (MULTI-EASE) and perform a task scheduling policy between multiple layers by formulating a delay-constrained energy minimization problem which minimizes the total energy consumption of the system subjected to the latency and emotion service quality constrains. The main contributions of this article are as follows:

\begin{figure*}
\centering
\includegraphics[width=5.5in]{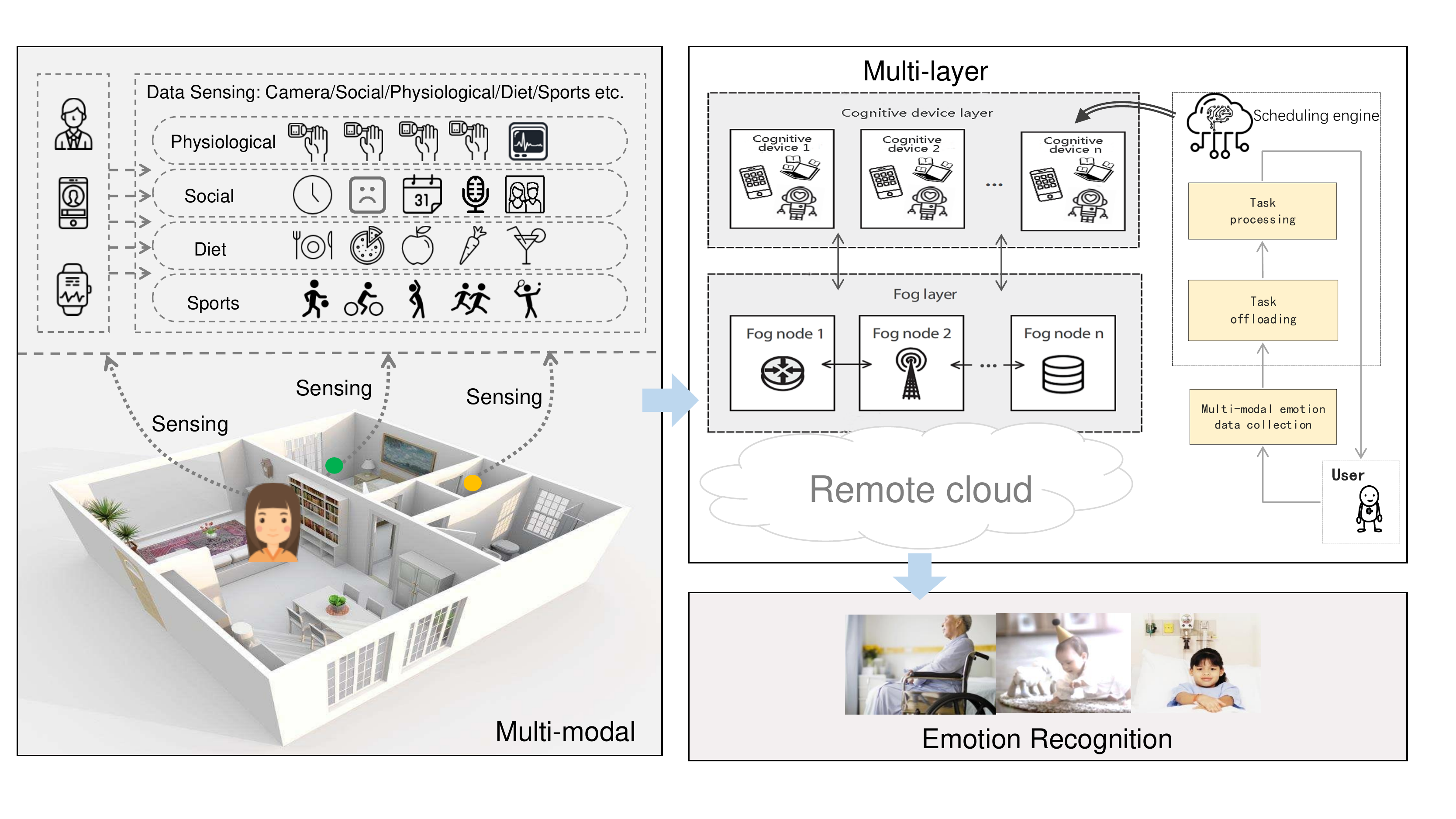}
\caption{MULTI-EASE Architecture}
\label{fig01}
\end{figure*}

\begin{enumerate}
\item We propose a MULTI-EASE architecture and perform a task scheduling policy between multiple layers to reduce the end-to-end delay and energy consumption by using an edge-based approach;
\item We formulate a delay-constrained energy minimization problem to improve the users' emotion interactive experience and achieve energy saving;
\item We deploy a prototype system to validate the architecture of MULTI-EASE that is proved to be a sustainable and efficient platform for emotion analysis applications.
\end{enumerate}

The remainder of this article is organized as follows. Section \ref{sec:arch} introduces the MULTI-EASE architecture. Section \ref{sec:model} introduces the modeling of multi-layer task scheduling and the analysis of delay and energy. Section \ref{sec:proto} presents the MULTI-EASE prototype. Section \ref{sec:performance} discusses the testing of main MULTI-EASE components. Finally, Section \ref{sec.conclusion} concludes the article.

\section{MULTI-EASE Architecture} \label{sec:arch}

Based on edge computing, this paper proposes a multi-modal multi-layer emotion-aware service (MULTI-EASE) architecture that considers multi-modal data as the data source of emotion recognition, and employs the intelligent terminal, edge server and cloud as multi-layer execution environment.
The architecture of MULTI-EASE is shown in Fig. \ref{fig01}.
The MULTI-EASE conducts a long-term intelligent and personalized emotion perception of a single user from multi-model data, and meanwhile performs a task scheduling policy between multiple layers to optimize the emotion recognition task.
Adaptive information fusion technologies ~\cite{tian17a} can also be integrated into MULTI-EASE in order to enhance the emotion perception accuracy.
Details of MULTI-EASE architecture are introduced as follows.

\subsection{Multi-modal}
At present, the emotion data sources are very abundant, as shown in Fig. \ref{fig01}, the data could be audio-visual information, physiological signal, diet and sports information, etc. In this paper, the facial and audio data are adopted to achieve emotion recognition, and facial emotion detection is characterized by higher recognition accuracy and lower computation overhead, while the audio emotion detection requires high accuracy of the original audio data. This paper mainly adopts the intelligent terminal such as an interactive robot to collect audio-visual data of the user and assist in better detection and recognition of user emotion.

\subsection{Multi-layer}
After acquiring the emotion data, a terminal always has to offload the computing tasks.
Networking between different edge terminals can be implemented based on some existing advanced energy-efficient communication mechanisms such as ~\cite{tian16r}.
The emotion data of users are personalized, and the demands on delay are strict for users, so that the emotion computing tasks may be offloaded to different service nodes for processing, such as other terminals, edge servers and cloud platforms as shown in Fig. \ref{fig01}.

\textbf{a) Terminal device}: In the emotion recognition, an intelligent device or emotion recognition robot denotes as a local terminal device. The terminal device has the function of collecting the emotion data. For instance, a robot can collect facial expression and audio of a user, and the smart clothing and other wearable devices can collect the physiological data of a user.
However, both computing and storage capability and the battery energy of terminal devices are limited.
When the number of users' requests is excessive, or the complexity of a computing task is larger than available computing power, it is not suitable to perform the task directly on a terminal device.
On the other hand, local devices are the closest devices to user, thus, they have low communication latency and are suitable for processing simple task with sensitive latency.

\textbf{b) Edge server}: A part of computing tasks can be processed on a network edge by introducing the edge computing to reduce network congestion. The edge computing layer is composed of many nodes possessing the computing capability. These edge nodes can be gateway, router, exchange or local server, etc. In emotion recognition, a local server near a terminal device is considered as an edge node. The computing capability of edge computing is weaker than that of cloud computing, but the communication latency of edge server is far below that of a cloud platform.
The performance of algorithm deployed on the edge server is generally lower than that of algorithm deployed on the cloud, but algorithm deployed on the edge server can rapidly feed-back the result to a user. The computing capability and storage capability of an edge computing node are also limited. Therefore, more complex computation should be executed on a remote cloud platform.

\textbf{c) Cloud platform}: Both local computing and edge computing may relieve network congestion and reduce communication latency, while cloud computing does not process unnecessary computing tasks and can focus on high-precision computation and analysis to provide an optimal computing service for users. The infrastructure of cloud computing is based on a cloud platform that deploys a high-performance emotion recognition algorithm.

\textbf{d) Scheduling engine}: By using the SDN (Software Defined Network, SDN) technology, we can deploy a scheduling engine in three layers of a network framework. The scheduling engine processes real-time multimodal emotional data flow in network environment, and can execute task scheduling policy through perceiving the computing resources, environmental communication resources, and network resources (such as network type, business data flow, communication quality, and other dynamic environment parameters) of a three-layer structure. The scheduling engine has global information about multi-layer execution environment and determines the scheduling policy on line according to current task requests and network resources.

\section{Modeling and Analysis} \label{sec:model}

\subsection{Task model}

For each task $Q_{i}$, it can be expressed as $Q_{i}=\{\omega_{i},s_{i},o_{i},a_{i}\}$. $\omega_{i}$ is the computation resource for task $Q_{i}$ and can be measured by the number of CPU cycles required to accomplish the task; $s_{i}$ specifies the size of the task in bits; $o_{i}$ represents the amount of data generated by the task feedback; $a_{i}$ represents the quality of task performance.
In emotion detection , $\omega_{i}$ is the computation resource required for the emotion detection task; $s_{i}$ is the emotion data (image, voice, video, etc.); $o_{i}$ is the amount of data fed back to the client; and $a_{i}$ is the recognition accuracy of emotion detection. In addition, $d_{i}$ represents the task duration of task $Q_{i}$. Considering that the output of a task is generally much smaller than the task size, and can be returned to the device with negligible transmission delay (e.g., face recognition and language processing), the feedback time can be represented by $\xi_{i}(o_{i})$ which is viewed as a constant and can be ignored.

The delay and energy consumption in different layers will be discussed below.

\subsection{Multi-layer task scheduling}

\subsubsection{Local Layer}

The latency of emotion computing task on a terminal device is given by (\ref{eq.eq1}). In (\ref{eq.eq1}), $T_{n,i}^{loc}$ is the time required to perform the task locally at the terminal device $n$.
We assume that the CPU at the $n$th local device is operating at frequency $f^{loc}_{n}$ (in Hz) if a task is being executed, and its energy consumption is given by $P_n^{loc}$ (in W); otherwise, the local CPU is idle and consumes no energy. The number of required CPU cycles for executing a task successfully is denoted as $\omega_{i}$, which depends on the types of mobile applications as introduced in the task definition $Q_{i}=\{\omega_{i},s_{i},o_{i},a_{i}\}$.
Then the task duration of local processing can be represented by $d_{n,i}^{loc}$, as given by


\begin{equation}
d_{n,i}^{loc}=T_{n,i}^{loc}=\frac{\omega_{i}}{f^{loc}_{n}}.
\label{eq.eq1}
\end{equation}

The CPU power consumption is widely modeled to be a superlinear function of $f^{loc}_{n}$, as given by $P_n^{loc}=\kappa(f^{loc}_{n})^\gamma$. $\kappa$ and $\gamma$ are pre-configured model parameters depending on the chip architecture. According to ~\cite{Wen2017Energy}, the energy consumption of device $n$ for local computation, denoted by $E^{loc}_{n,i}$, is therefore given by

\begin{equation}
E^{loc}_{n,i}=P_n^{loc} \cdot T_{n,i}^{loc} = \kappa(f^{loc}_{n})^{\gamma -1}{\omega_{i}}
\end{equation}

Typically, $\kappa = 10^{-26}$ denotes the energy coefficient that depends on the chip architecture~\cite{zhang13e}, and $2 \le \gamma \le 3$ ~\cite{chen16e}.

\subsubsection{Edge Layer}

In order to offload a computation task to the edge server, all the input data of the task should be successfully delivered to the edge server over the wireless channel.
We assume $P_{n}^{tr}$ is the uplink transmission power of terminal device $n$, $h_{n,m}$ denotes the channel power gain between the device $n$ and the edge server $m$, $B$ is the system bandwidth and $N_0$ is the noise power spectral density at the receiver; According to the work in ~\cite{b16e}, the uplink data rate $r_{n,m}$ of device $n$ can be given by

\begin{equation}
r_{n,m} = Blog_2(1+\frac{P_{n}^{tr}h_{n,m}}{N_0})
\end{equation}

The latency of edge computing is given by (\ref{eq.eq2}) where $T_{n,m}^{loc \to edge}$ is the time needed to offload emotion data to a edge server $m$. $T_{m,i}^{edge}$ is the execution time of task $i$ on edge server $m$, $T_{m,n}^{down}$ is the time needed to feedback the computation results to terminal device $n$. Assume that $f^{edge}_{m}$ represents the computing capacity of the $m$th edge node, then the task duration on edge node can be represented by $d^{edge}_{m,i}$.


\begin{equation}
d_{i}^{edge}=T_{n,m}^{loc \to edge}+T_{m,i}^{edge}+T_{m,n}^{down}=\frac{s_{n,m}}{r_{n,m}}+\frac{\omega_{i}}{f^{edge}_{m}}+\xi_{i}(o_{i}).
\label{eq.eq2}
\end{equation}


The energy consumption can be calculated as the communication consumption between the local device and edge node. Let $P_{n}^{tr}$ denotes the transmission power of terminal device $n$ and $P_{n}^{id}$ denotes the idle power.
When the task is executed on the edge server, the mobile device needs to wait for the return of the response result. The idle power consumption of the mobile device can be calculated as $P_{n}^{id}T_{m,i}^{edge}$.
Then, the energy consumption is defined as the sum of idle power consumption and data transmission power consumption which can be expressed as $E^{edge}_{m,i}$ as follows:

\begin{equation}\label{eq1}
E^{edge}_{m,i}=P_{n}^{tr}T_{n,m}^{loc \to edge}+P_{n}^{id}T_{m,i}^{edge}=P_{n}^{tr}\frac{s_{n,m}}{r_{n,m}}+P_{n}^{id}\frac{\omega_{i}}{f^{edge}_{m}}
\end{equation}

\subsubsection{Cloud Layer}
\begin{figure*}
\centering
\includegraphics[width=5.5in]{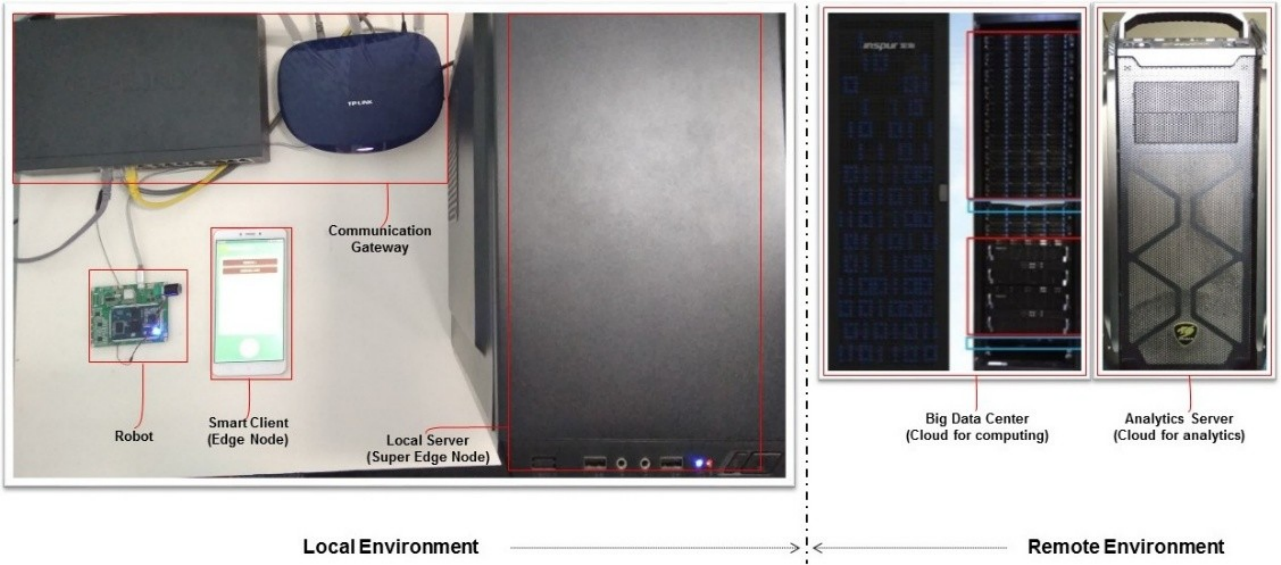}
\caption{MULTI-EASE Prototype}
\label{fig02}
\end{figure*}
The latency of cloud computing is given by (\ref{eq.eq3}), where $T_{n,k}^{loc \to cloud}$ is the time for emotion data offloading to the remote cloud server $k$, and $T_{k,i}^{cloud}$ is the execution time of task $i$ on cloud server $k$. Assume that $f^{cloud}_{k}$ represents the computing capacity of the $k$th cloud server, then the task duration on cloud can be represented by $d^{cloud}_{k,i}$.

\begin{equation}
d_{k,i}^{cloud}=T_{n,k}^{loc \to cloud}+T_{k,i}^{cloud}+T_{k,n}^{down}=\frac{s_{n,k}}{r_{n,k}}+\frac{\omega_{i}}{f^{cloud}_{k}}+\xi_{i}(o_{i}).
\label{eq.eq3}
\end{equation}

The energy consumption can be calculated as the communication consumption between the local device and cloud, as the same case of edge computing, the energy consumption is defined by $E^{cloud}_{k,i}$ as follows:

\begin{equation}\label{eq1}
E^{cloud}_{k,i}=P_{n}^{tr}T_{n,k}^{loc \to cloud}+P_{n}^{id}T_{k,i}^{cloud}=P_{n}^{tr}\frac{s_{n,k}}{r_{n,k}}+P_{n}^{id}\frac{\omega_{i}}{f^{cloud}_{k}}
\end{equation}


\subsection{Delay and Power Analysis}
In this subsection, we will analyze the average delay of each task and the average power consumption at the mobile device by modeling the MULTI-EASE system.

Let $p_{i}^{loc}$,$p_{i}^{edge}$,$p_{i}^{cloud}$ represent the offloading policy of task $Q_{i}$. If $ Q_{i}$ is processed by a local device $n$, then $p_{n,i}^{loc}$ equals to 1, otherwise it equals to 0. Similarly, if $Q_{i}$ is processed by a edge $m$, then $p_{m,i}^{edge}$ equals to 1, otherwise it equals to 0. If $Q_{i}$ is processed by a cloud $k$ then $p_{k,i}^{cloud}$ is 1, otherwise it equals to 0. Also, the sum of $p_{i}^{loc}$£¬$p_{i}^{edge}$£¬$p_{i}^{cloud}$ should equal to 1 as defined by (11).

\begin{eqnarray}p_{n,i}^{loc}=
\begin{cases}\label{eq8}
1, &\mbox{if task $i$ is processed by local device $n$,}\cr 0, &\mbox{otherwise.} \end{cases}
\end{eqnarray}

\begin{eqnarray}p_{m,i}^{edge}=
\begin{cases}\label{eq9}
1, &\mbox{if task $i$ is processed by edge $m$,}\cr 0, &\mbox{otherwise.} \end{cases}
\end{eqnarray}

\begin{eqnarray}p_{k,i}^{cloud}=
\begin{cases}\label{eq10}
1, &\mbox{if task $i$ is processed by cloud $k$,}\cr 0, &\mbox{otherwise.} \end{cases}
\end{eqnarray}

\begin{equation}\label{eq11}
p_{n,i}^{loc}+p_{m,i}^{edge}+p_{k,i}^{cloud}=1.
\end{equation}

Then, the task duration $d_{i}$ and energy consumption $E_{i}$ of task $Q_{i}$ can be expressed by:

\begin{equation}\label{eq12}
d_{i}=p_{n,i}^{loc}d_{n,i}^{loc}+p_{m,i}^{edge}d_{m,i}^{edge}+p_{k,i}^{cloud}d_{k,i}^{cloud}.
\end{equation}

\begin{equation}\label{eq13}
E_{i}=p_{n,i}^{loc}E_{n,i}^{loc}+p_{m,i}^{edge}E_{m,i}^{edge}+p_{k,i}^{cloud}E_{k,i}^{cloud}.
\end{equation}

Considering the delay, energy consumption and quality of service are different at different layer levels, and the user's sensitivity to the delay of emotion recognition is also different.
We formulate a delay-constrained energy minimization problem to minimize the energy consumption while satisfying user delay requirements and accuracy of algorithm.
Let assume user's requirement be defined as $U_{i}=\{A_{i}, D_{i}\}$, where $A_{i}$  represents the requirement for task performance (accuracy in emotion detection), and $D_{i}$ is the task deadline, i.e., the maximum delay that the task can tolerate.

\begin{equation}\label{eq14}
d_{i} \leq D_{i}.
\end{equation}

\begin{equation}\label{eq15}
a_{i} \geq A_{i}.
\end{equation}


Assume that the total number of tasks initiated within the service area is $q$. An energy efficiency optimization problem, which is to minimize the overall energy consumption while satisfying user's requirement, can be formulated by:

\begin{equation}\label{eq16}
\begin{aligned}
& {\text{minimize:}}
& & \frac{1}{q}\sum\limits_{i=1}^q E_i,\\
& \text{subject to:}
& & (2)-(15). \\
\end{aligned}
\end{equation}

The problem defined by (16),
is a typical 0-1 integer linear optimization problem. By the use of branch and bound algorithm~\cite{convex}, the optimal solution can be obtained. Thus, the linear iterative algorithm can be utilized and obtain the approximate optimal solution.

\section{MULTI-EASE test platform } \label{sec:proto}

\subsection{Prototype system}
The MULTI-EASE prototype platform is shown in Fig. \ref{fig02}, where it can be seen that intelligent terminal and local server are two typical edge computing nodes, cloud platform is a data center and GPU server is an analysis server.

\begin{table*}
  \caption{MULTI-EASE Prototype Platform Profile}
  \begin{center}
  \begin{tabular}{|c|c|c|c|}
    \hline
     \multicolumn{4}{|c|}{\textbf{MULTI-EASE Hardware Parameters}} \\
    \hline
    \textbf{Node}     & \multicolumn{2}{|c|}{\textbf{Core Processor}} & \textbf{Memory} \\    
    \hline
    \textbf{Robot}           & \multicolumn{2}{|c|}{4 x ARM Cortex-A33, 1.2GHz}      & 1GB LPDDR2 \\
    \hline
    \textbf{Edge Server}    & \multicolumn{2}{|c|}{Intel Cor 2 Quad 9400, 2.66GHz}  & 8GB DIMM \\
    \hline
    \textbf{Cloud}           & \multicolumn{2}{|c|}{AMD FX 8-Core, 4GHz}             & 32 GB RAM DDR3 \\
    \hline

     \multicolumn{4}{|c|}{\textbf{Experimental Parameters Settings}} \\
    \hline
                             & \textbf{Robot} & \textbf{Edge Server}   &  \textbf{Cloud}  \\    
    \hline
    \textbf{Algorithm}       &  Facial + Voice     & Facial + Voice & Facial + Voice \\
    \hline
    \textbf{Concurrent Invocation(Facial)}    & 1 to 60  & 50 to 550  & 50 to 700\\
    \hline
    \textbf{Concurrent Invocation(Voice)}    & 1 to 35  &  50 to 550 &  50 to 700\\
    \hline

  \end{tabular}
  \label{tab.01}
  \end{center}
\end{table*}

\begin{figure*}
\centering
\includegraphics[width=1.3\columnwidth]{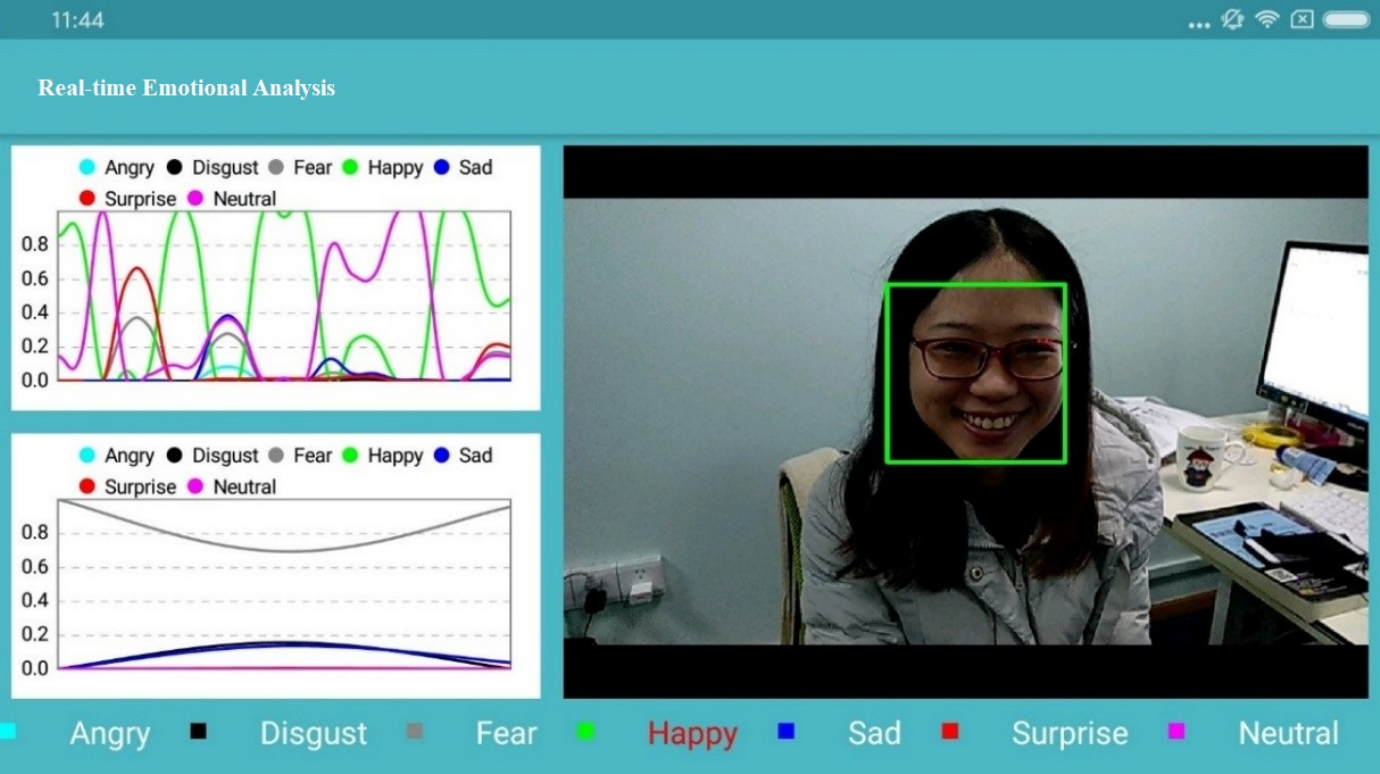}
\caption{Emotion Recognition Demo}
\label{fig03}
\end{figure*}

The MULTI-EASE hardware parameters presented in Table \ref{tab.01} show the hardware profile of MULTI-EASE prototype platform and list the computing performance of a device under three different levels (terminal device, local server (edge node), and cloud platform). Each MULTI-EASE layer has different computing capability and different working load distributed to it. While generating the working load, different numbers of service requests are generated following the uniform probability distribution, and different load scenes are simulated. ``Experimental parameters settings'' in Table \ref{tab.01} shows the statistics of test executed on devices at each level.
We deploy facial expression recognition and voice emotion recognition algorithms on each layer. The concurrent request time is set from 1 to 60 at Robot for facial recognition, and from 1 to 35 at Robot for voice recognition, that is because voice recognition task consumes more computation resources. Cloud has higher load capacity than local server so the concurrent invocation cloud be from 50 to 700 while it is from 50 to 550 at local server. In the experiment, we conduct several stress tests on multi-layers based on above settings.

\subsection{Emotion Analysis}

The emotion recognition algorithms used in this paper are facial and voice emotion recognition. The model training was executed based on an independent GPU algorithm server. At the end of the training, the trained model was saved in the binary file in the form of ``.pb'' file. The model could be executed on a cloud platform and migrated to the terminal device or other devices for execution. The model execution on the edge device needed a TensorFlow environment, and the model execution on the terminal device needed a TensorFlow support package. Figure. \ref{fig03} shows the execution results of real-time emotion recognition in the terminal device. Two figures on the left side of Fig. \ref{fig03} show the probability distribution of results of image emotion recognition (upper figure) and voice emotion recognition (lower figure). On the interface bottom, seven specific emotions (angry, disgust, fear, happy, sad, surprise, and neutral) are shown. The emotion mark colored in red is currently recognized emotion. The image used in the test was captured by a self-contained camera of a terminal device. The OpenCV can catch the human face in the image, conduct the corresponding preprocessing, and transmit obtained result to the model as an input data. The audio data were collected by a microphone (MIC), and the collected voice data denoted an input to the emotion model for real-time recognition. The specific algorithms are introduced as below.

\begin{itemize}
  \item \textbf{Facial emotion recognition:} The facial emotion recognition firstly realizes a facial expression captured by OpenCV, then realizes a face alignment by the MTCNN (Multi-task Cascaded Convolutional Networks)~\cite{MCNN}, and finally realizes the facial feature training by the VGGNET to generate a 256-dimension facial feature vector for executing the task of expression classification.
  \item \textbf{Voice emotion recognition:} First, the speech framing is conducted, then the MFCC is adopted for feature extraction, the static speech feature vector is generated, the first and second derivatives of a generated static speech feature are evaluated, and a three-channel speech spectrogram is obtained. Finally, the AlexNet DCNN (deep convolutional neural network) model for emotion feature extraction is adopted to generate 512-dimension emotion feature vector for executing the task of speech emotion classification \cite{2}.
\end{itemize}

\section{Performance evaluation of emotion recognition } \label{sec:performance}

We introduce three baseline task scheduling policies, including the \textbf{local execution policy}, which executes all the computation tasks locally at the mobile device; the \textbf{cloud execution policy}, where all the tasks are offloaded to the cloud for computing; and \textbf{random execution policy}, which randomly select one layer for task execution. Our proposed task scheduling policy called \textbf{edge-based multi-layer execution policy} solves the delay-constrained energy minimization problem to reduce the end-to-end delay and energy consumption towards specific latency and energy demands of emotion analysis task.


For large scale performance analysis and evaluation, the numerical simulations are designed except for the real prototype system. The simulations are deployed based on real-world settings according to our real prototype system. All the parameters, including the energy consumption rates and computing capacity, are measured from real mobile devices.
In the experimental setup, we assume that system bandwidth $B$ is 1 MHz, the local processing power $P_n^{loc}$, the standby power $P_n^{id}$, and the transmitting power $P_{n}^{tr}$ of mobile device are 0.9 W, 0.3 W and 1.3 W respectively.
The corresponding Gaussian channel noise $N_0$ and channel power gain $h_{n,m}$ are $10^{-9}$ W and $10^{-5}$, respectively. For task $Q_i$, we assume that required computing capacity $w_i$ and data size $s_i$ are generated by a probability distribution~\cite{tang16a}.
Furthermore, we assume that the computing capabilities of cloud server $f_n^{cloud}$, edge server $f_n^{edge}$ and mobile device $f_n^{loc}$ are 15GHz, 10 GHz and 2 GHz, respectively.

Below we discuss the task computing time at different layers, and the analysis of task delay and energy consumption. For convenience, we illustrate the performance indexes in detail in Table \ref{tab.02}.

\begin{table*}
  \renewcommand{\arraystretch}{1.3}
  \caption{Performance Indexes Illustration}
  \begin{center}
  \begin{tabular}{l|l|l}
    \hline
    \textbf{Indexes} &  \textbf{Symbols}  & \textbf{Illustration} \\
    \hline
    \multirow{3}{*}{Task computing time} & $T^{loc}$  & Task computing time at local layer \\
        \cline{2-3}
        &  $T^{edge}$  & Task computing time at edge layer\\
        \cline{2-3}
        &  $T^{cloud}$  & Task computing time at cloud layer\\
    \hline
    Task delay & $\sum_{i=1}^qd_i$ &  $d_i$ is the task duration of task $Q_{i}$  defined in (12)\\
    \hline
    Task energy consumption & $\sum_{i=1}^qE_i$ & $E_i$ is the task energy consumption of task $Q_{i}$  defined in (13) \\
    \hline

  \end{tabular}
  \label{tab.02}
  \end{center}
\end{table*}

\subsection{Stress testing for task computing time}

\begin{figure}
\centering
\subfigure[Task computing time for facial emotion recognition]{\label{fig05.a} \includegraphics[width=\columnwidth]{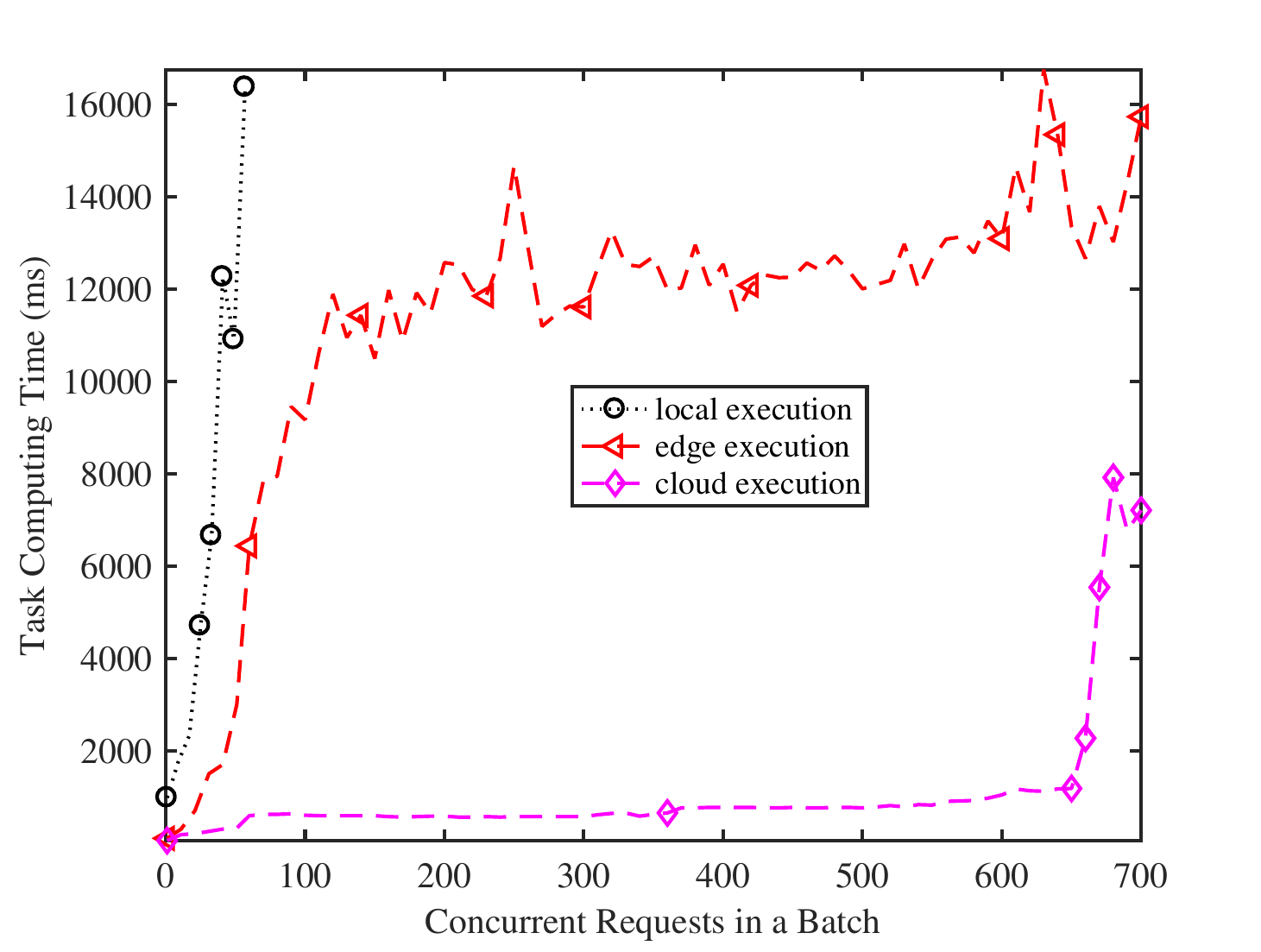}}
\subfigure[Task computing time for voice emotion recognition]{\label{fig05.b} \includegraphics[width=\columnwidth]{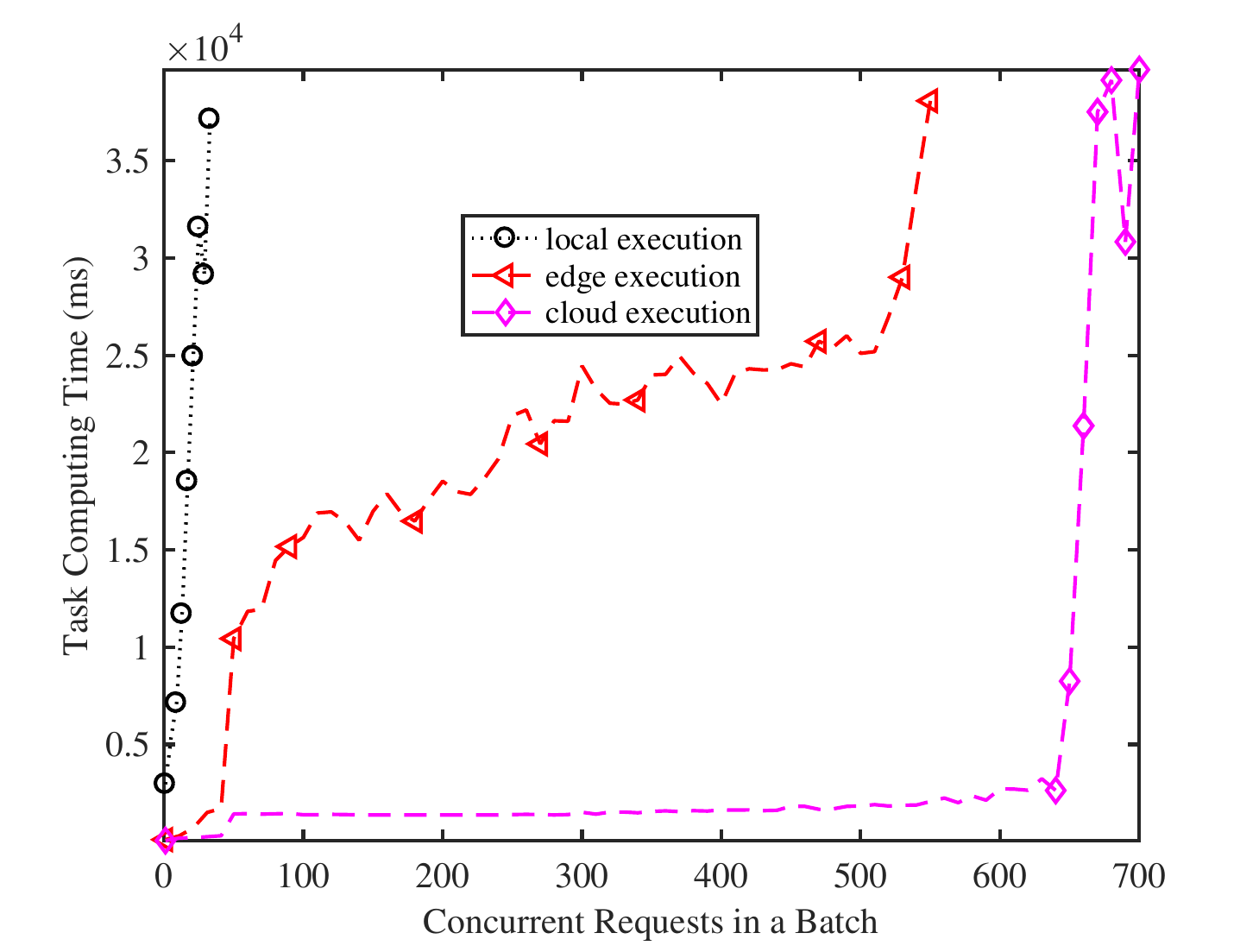}}
\caption{Stress testing for task computing time at different Layers}
\label{fig.stress_d}
\end{figure}

\begin{figure*}
\centering
\subfigure[Task delay with task arrival number $q$]{\label{fig.scale.a} \includegraphics[width=\columnwidth]{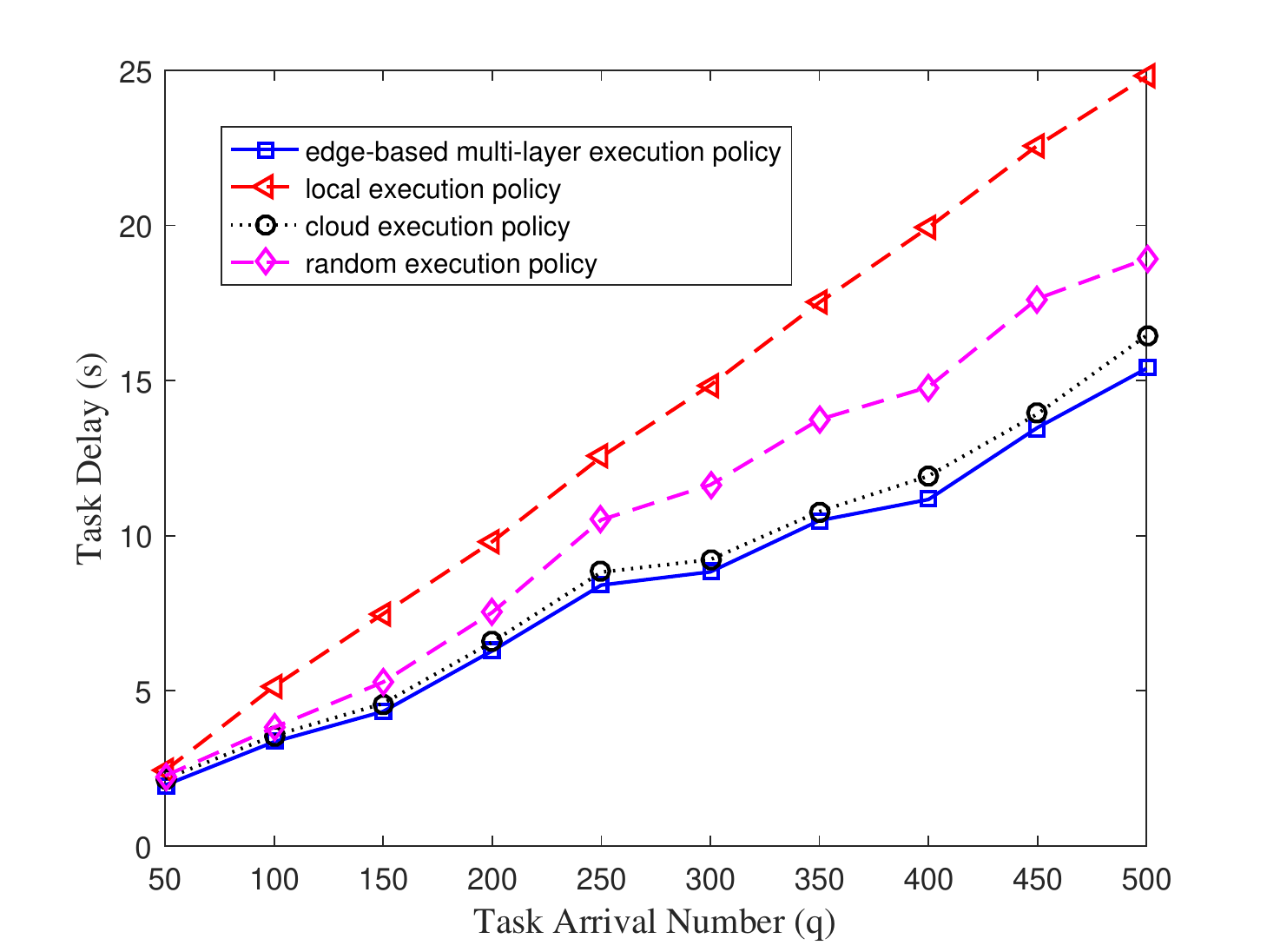}}
\subfigure[Energy consumption with task arrival number $q$, where s=10 MB]{\label{fig.scale.b} \includegraphics[width=\columnwidth]{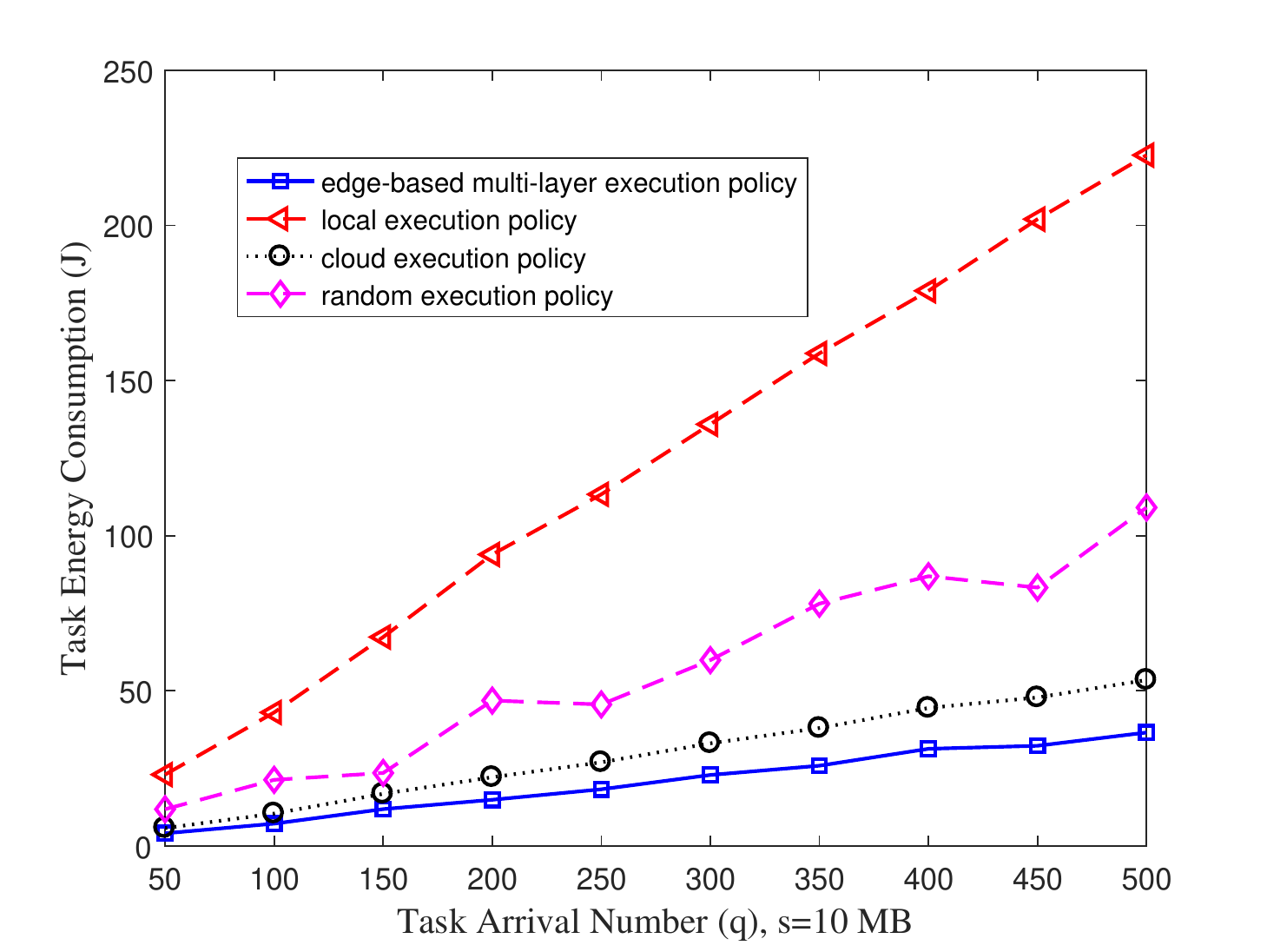}}
\subfigure[Energy consumption with task arrival number $q$, where s=110 MB]{\label{fig.scale.c} \includegraphics[width=\columnwidth]{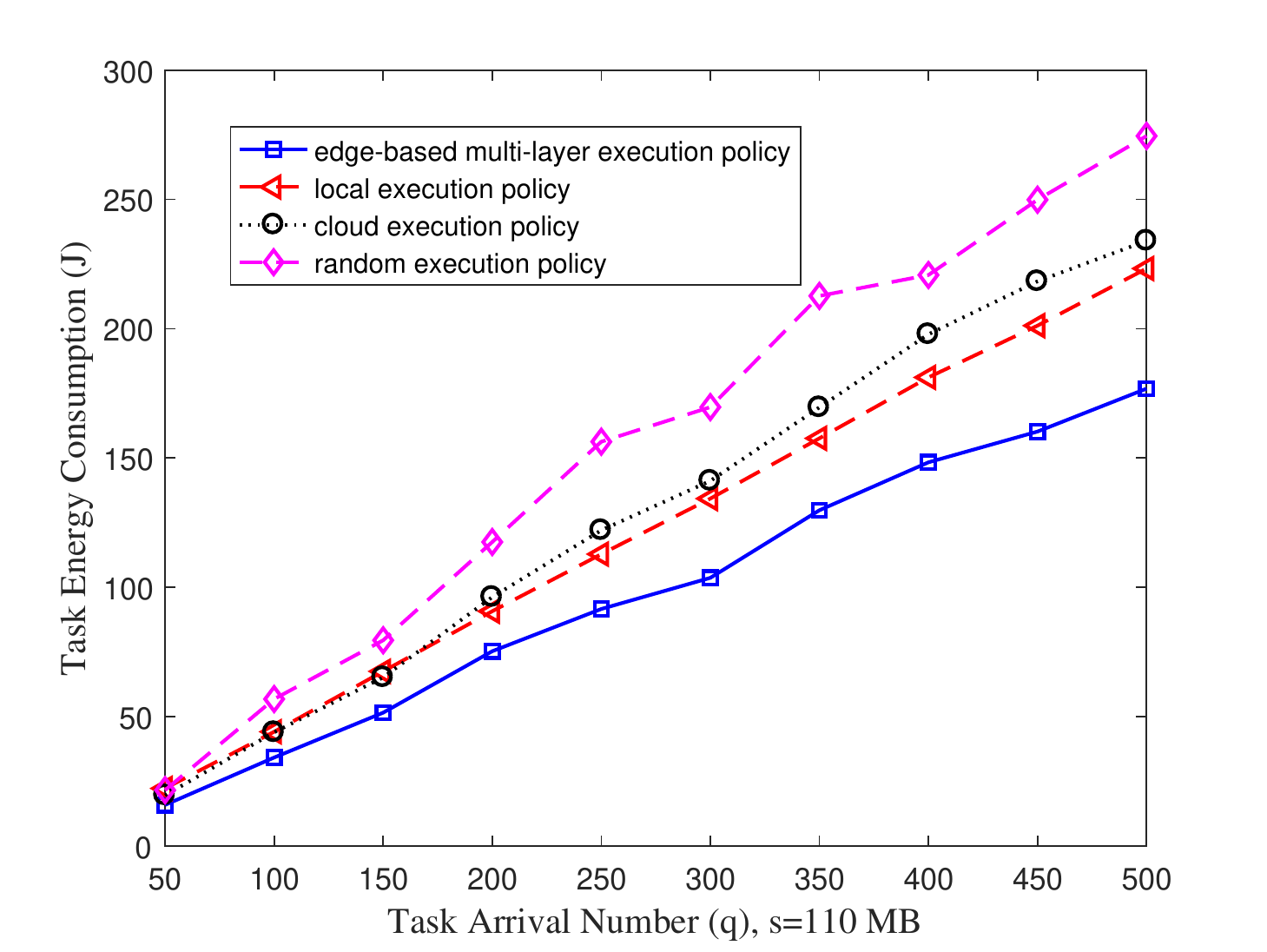}}
\caption{Delay and energy analysis with task scale}
\label{fig.scale}
\end{figure*}

The performance evaluation of emotion recognition algorithm using the MULTI-EASE system is based on facial emotion and voice emotion analysis.
We deploy facial expression recognition and voice emotion recognition algorithms on each layer.
The stress testing for task computing time at different layers can be seen from Fig. \ref{fig.stress_d}.

It is obvious that the local computing time increases sharply when the concurrent request number increases to 60 for facial recognition, and to 35 for voice recognition, that is due to the limited computational capabilities of local device, and voice recognition task consumes more computation resources which results in that it cannot response much concurrent requests as facial recognition task.
When the concurrent request number increases to 550 and 700 respectively, the edge computing time and cloud computing time increase sharply. Cloud has higher load capacity than local server so the concurrent invocation cloud be from 50 to 700 while it is from 50 to 550 at local server.
It can also be seen that voice recognition task needs much more computing time for its higher computation demands.

\subsection{Delay and energy analysis with task scale}

The average delay is shown in Fig. \ref{fig.scale.a}, it can be observed from the figure that, the average delays achieved by the local execution, cloud execution, random execution and the edge-based multi-layer execution, increase with the average computation task arrival number $q$, which is in accordance with our intuition.
The task delay of local execution policy is much larger than that of other policies, this is due to the fact that the execution time required by cloud server and edge server is much smaller than that by the local CPU.
When $q$ increases, more arrived tasks should be sent to the edge server or cloud server for lower latency.

In the energy efficiency performance aspects, simulations results show the superiority of the proposed scheme.
Fig. ~\ref{fig.scale.b} and Fig. ~\ref{fig.scale.c} show the energy consumption of the mobile device when the number of computation task increases from 50 to 500 with four different task offloading policies, considering two cases with different task data size, i.e., $s_i=10 MB$ and $s_i=110 MB$.
It is obvious that the energy consumption of our proposed policy is the lowest in both two cases compared with other three policies. However, the gap of energy consumption between local execution policy and cloud execution policy is different. It can be seen that the energy consumption of local execution policy is much larger than that of cloud execution policy when the task data size is small, while it is almost the same and even lower than that of cloud execution policy when the task data size is large. This is because it results in large transmission latency and energy when the task data size is too large.

\subsection{Performance analysis on energy consumption}

\begin{figure*}
\centering
\subfigure[Average energy consumption as D req increases from 1s to 3.5s.]{\label{fig.energy.a} \includegraphics[width=\columnwidth]{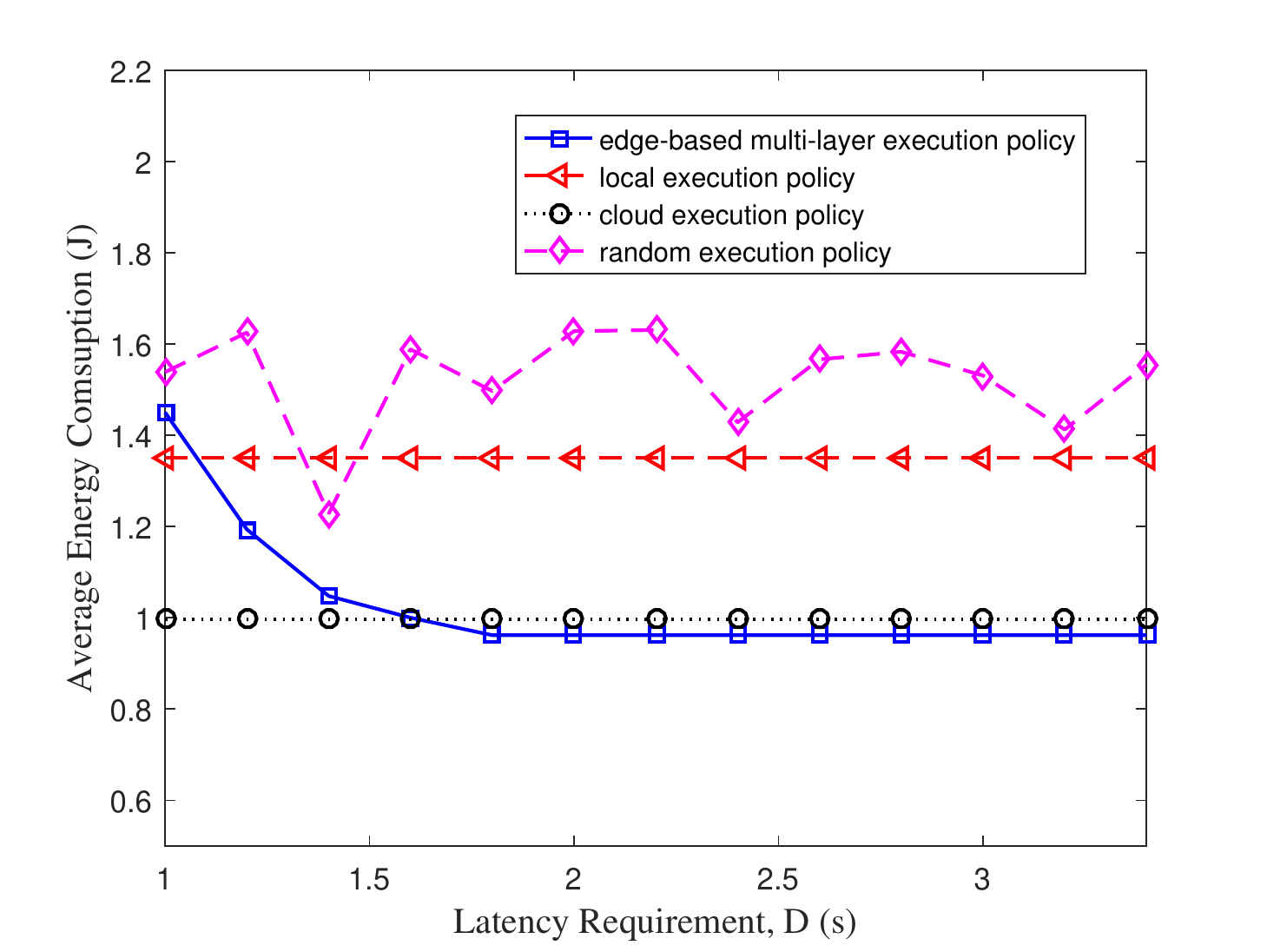}}
\subfigure[Average energy consumption as s increases from 10 to 80 MB.]{\label{fig.energy.b} \includegraphics[width=\columnwidth]{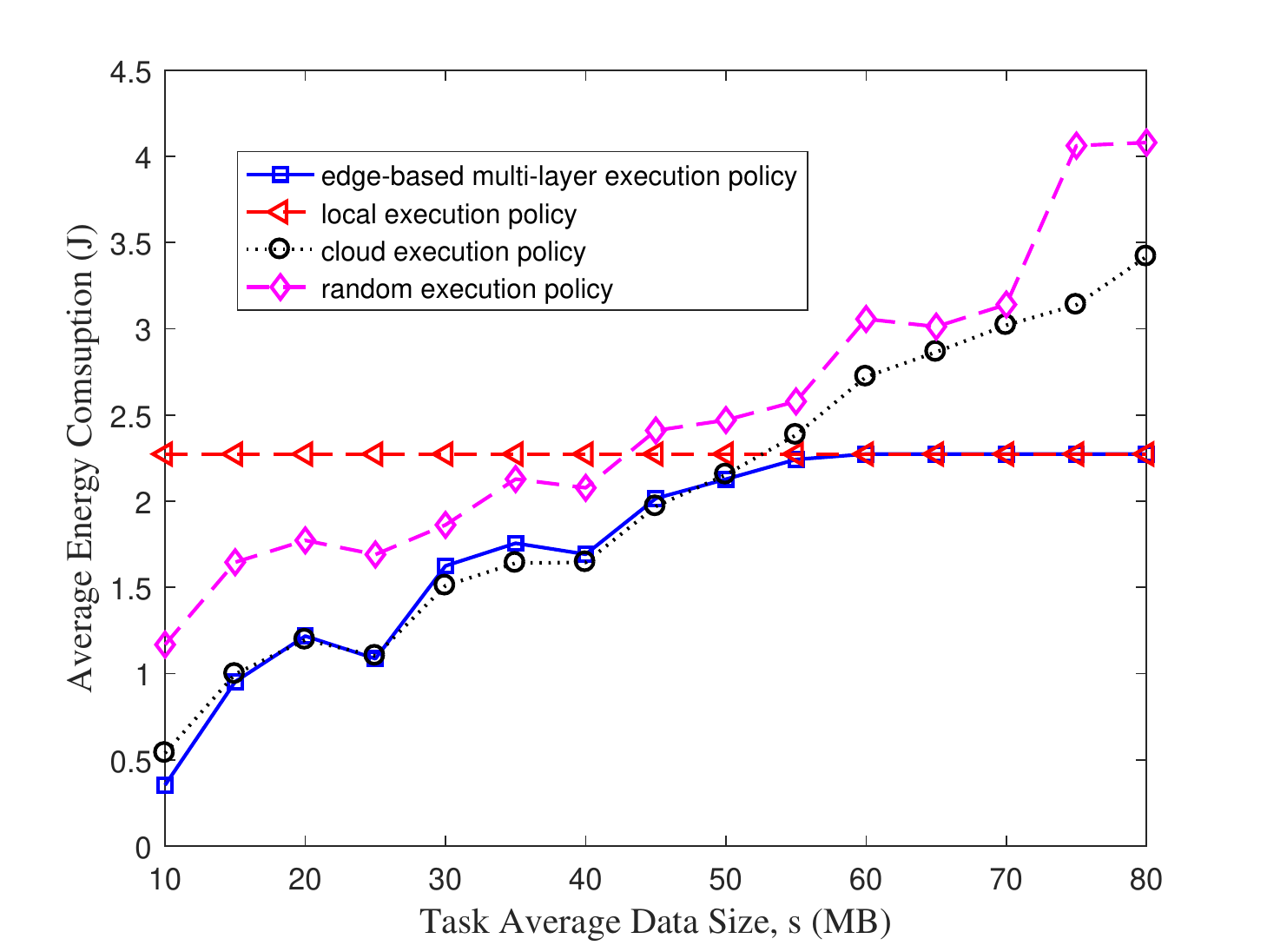}}
\subfigure[Average energy consumption as w increases from 0.1 to 10 gigacycles.]{\label{fig.energy.c} \includegraphics[width=\columnwidth]{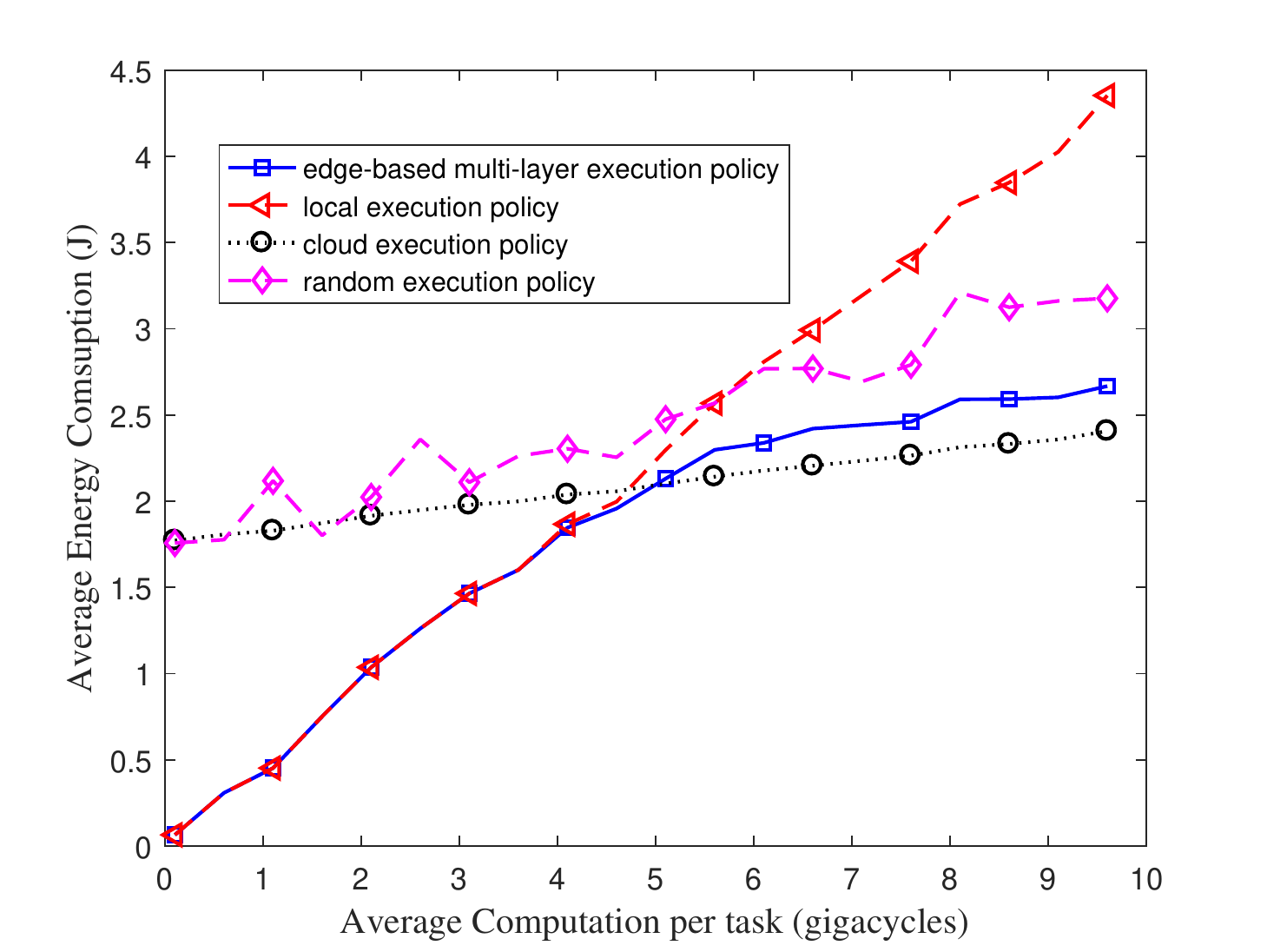}}
\caption{Performance analysis on energy consumption with different influence factors.}
\label{fig.energy}
\end{figure*}

In this subsection, we will discuss the influence of different factors on energy consumption, i.e., the task deadline, the task data size and the task computation amount. Obviously, the proposed method can always find a better energy-saving solution than other three approaches with the changes of different factors as shown in Fig. \ref{fig.energy}.

Fig. \ref{fig.energy.a} shows the average energy consumptions when task deadline D ranges from 1s to 3.5s. With the growth of D, the energy consumptions of edge-based multi-layer execution policy first decrease and then stabilize at 0.95 J when $D\ge 1.6s$. This is because increasingly relaxed deadlines allow a growing number of devices to be selected for offloading, thereby reducing energy consumption. When the deadlines are so loose that almost all tasks can be selected for offloading, it can be seen that our policy has lower energy consumption compared with cloud execution policy when deadlines are loose.

Fig. \ref{fig.energy.b} shows the average energy consumptions when task data size $s$ ranges from 10 to 80 MB. With the growth of $s$, the energy consumptions of edge-based multi-layer execution policy first increase along with that of cloud execution policy, and then stabilize at 2.3 J when $s\ge 55MB$. It indicates that in our proposed scheduling policy, the task is offloaded to edge server or cloud server when the task data size is small and is executed locally when the data size is larger than 55 MB.

Fig. \ref{fig.energy.c} shows the average energy consumptions when task computation amount $s$ ranges from 0.1 to 10 gigacycles. With the growth of $w$, the energy consumptions of edge-based multi-layer execution policy first increase along with that of local execution policy, and then slowly increase when $w \ge 5 gigacycles$.
When the task computation amount is small, the task should better be processed locally for lower energy consumption caused by data transmission. In contrast, when $w \ge 5 gigacycles$, the energy consumption of cloud execution is smaller due to its large computational capabilities resulting in lower latency and lower idle energy consumption for terminal device.
The average energy consumption achieved by the random execution policy fluctuates in all cases since it randomly selects one layer for task execution.



\section{Conclusion}\label{sec.conclusion}

In this article, the MULTI-EASE architecture that could reduce the end-to-end delay and energy consumption by using an edge-based approach is presented. By analyzing the average delay of each task and the average power consumption at the mobile device, we formulated a delay-constrained energy minimization problem which minimizes the total energy consumption of the system subjected to the latency and emotion service quality constrains. 
The multi-layer solution can find a kind of compromise scheduling scheme in terms of processing power availability than single-layer at device, edge server and cloud, by considering that the delay, energy consumption and quality of service are different at different layer levels, and the user's sensitivity to the delay of emotion recognition is also different.
A prototype system is also implemented to validate the architecture of MULTI-EASE to be a sustainable and efficient platform for emotion analysis applications by comparing it with systems not using such edge-based scheme. It is shown that the multi-layer solution always find a better energy-saving solution than other three approaches like local execution policy, cloud execution policy and random execution policy.

However, in this paper, we did not consider the storage capability of multi-layer execution environment, because for the emotional recognition application, the recognition model is trained offline, only real-time emotional data are needed for processing. So for small scale concurrent requests, different layers are able to store the data needed to be processed. But when the request numbers are very huge and high amount of data need to be processed, we still need to consider the case, which would be our future work that will consider the storage capability.

\bibliographystyle{IEEEtran}

\begin{IEEEbiography}[{\includegraphics[width=1in,height=1.25in,clip,keepaspectratio]{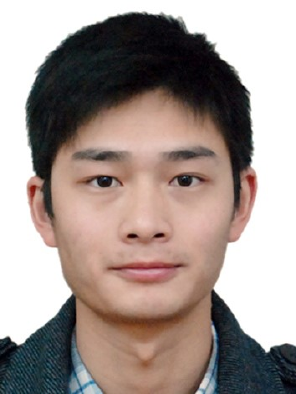}}]{Long Hu}
is a Ph.D student in School of Computer Science and Technology at Huazhong University of Science and Technology (HUST). He has also received his Master and B.S. degree in HUST. He is the Publication Chair for 4th International Conference on Cloud Computing (CloudComp 2013). Currently, his research includes 5G Mobile Communication System, Big Data Mining, Marine-Ship Communication, Internet of Things, and Multimedia Transmission over Wireless Network, etc.
\end{IEEEbiography}

\begin{IEEEbiography}[{\includegraphics[width=1in,height=1.25in,clip,keepaspectratio]{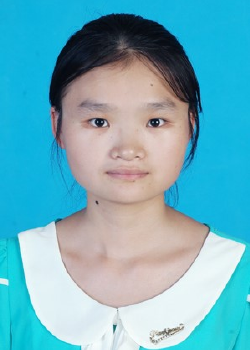}}]{Wei Li}
graduated from Wuhan University of Technology (WHUT) in 2015. Now, she is a Ph.D. candidate in Embedded and Pervasive Computing Lab of Huazhong University of Science and Technology (HUST), China. Her further research includes Pervasive Computing, The Internet of things, edge computing, cognitive computing, etc.
\end{IEEEbiography}

\begin{IEEEbiography}[{\includegraphics[width=1in,height=1.25in,clip,keepaspectratio]{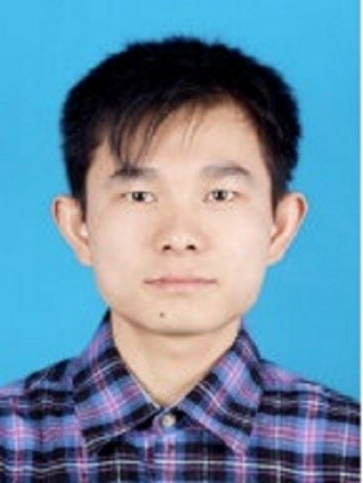}}]{Jun Yang}
received Banchelor and Master degree in Software Engineering from HUST, China in 2008 and 2011, respectively. Currently, he is a Ph.D candidate at Embedded and Pervasive Computing (EPIC) Lab in School of Computer Science and Technology, HUST. His research interests include cognitive computing, software intelligence, Internet of Things, cloud computing and big data analytics, etc.
\end{IEEEbiography}

\begin{IEEEbiography}[{\includegraphics[width=1in,height=1.25in,clip,keepaspectratio]{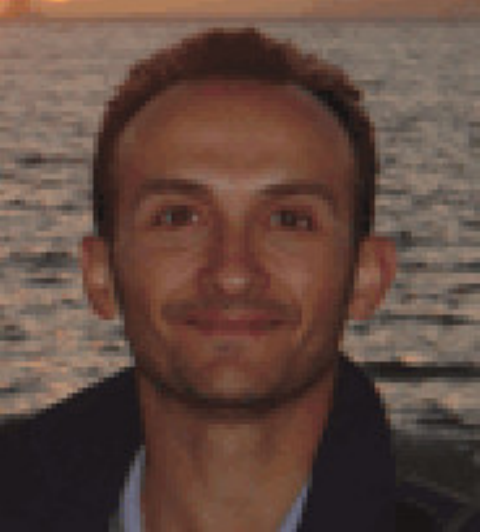}}]{Giancarlo Fortino}
(SM¡¯12) is Full Professor of Computer Engineering at the Dept. of Informatics, Modeling, Electronics, and Systems of the University of Calabria (Unical), Italy. He received a Ph.D. in Computer Engineering from Unical, in 1995 and 2000, respectively. He is also guest professor at Wuhan University of Technology (Wuhan, China), high-end expert at HUST (China), and senior research fellow at the Italian National Research Council ICAR Institute. He is the director of the SPEME lab at Unical as well as co-chair of Joint labs on IoT established between Unical and WUT and SMU Chinese universities, respectively. His research interests include agent-based computing, wireless (body) sensor networks, and Internet of Things. He is author of over 400 papers in int¡¯l journals, conferences and books. He is (founding) series editor of IEEE Press Book Series on Human-Machine Systems and EiC of Springer Internet of Things series and AE of many int'l journals such as IEEE TAC, IEEE THMS, IEEE IoTJ, IEEE SJ, IEEE SMCM, Information Fusion, JNCA, EAAI, etc. He is cofounder and CEO of SenSysCal S.r.l., a Unical spinoff focused on innovative IoT systems. Fortino is currently member of the IEEE SMCS BoG and of the IEEE Press BoG, and chair of the IEEE SMCS Italian Chapter.
\end{IEEEbiography}

\begin{IEEEbiography}[{\includegraphics[width=1in,height=1.25in,clip,keepaspectratio]{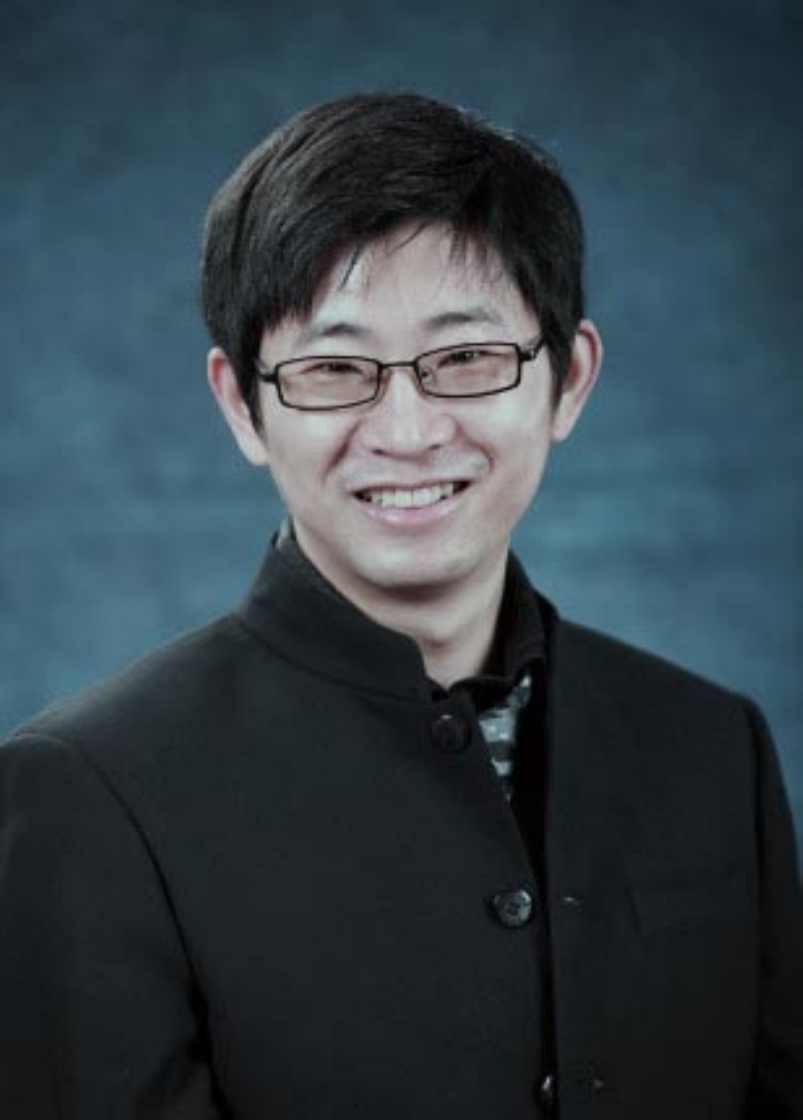}}]{Min Chen}
(M¡¯08¨CSM¡¯09) was an Assistant Professor with the School of Computer Science and Engineering, Seoul National University (SNU), from 2009 to 2012. He was a Post-Doctoral Fellow with SNU. He was a PostDoctoral Fellow with the Department of Electrical and Computer Engineering, University of British Columbia. He is currently a Professor with the School of Computer Science and Technology, Huazhong University of Science and Technology (HUST). He is also the Director of the Embedded and Pervasive Computing Laboratory. He has authored over 260 paper publications, including over 120 SCI papers, over 50 IEEE Transactions/Journal papers, six ISI highly cited papers and one hot paper. He has authored a book on IoT, OPNET IoT Simulation (HUST Press, 2015), a book on 5G,
Software Defined 5G Networks (HUST Press, 2016) and a book on big data, Big Data Related Technologies (2014) with Springer Series in Computer Science. His Google Scholars Citations reached over 8200 with an h-index of 45. His top paper was cited over 900 times. His research focuses on Internet of Things, mobile cloud, body area networks, emotion-aware computing, healthcare big data, cyber physical systems, and robotics. He received the Best Paper Award from the IEEE ICC 2012 and the Best Paper Runner-up Award from QShine 2008. He is currently a Guest Editor of the IEEE NETWORK and the IEEE Wireless Communications Magazine. He is the Co-Chair of the IEEE ICC 2012-Communications Theory Symposium and the Co-Chair of the IEEE ICC 2013-Wireless Networks Symposium. He is the General Co-Chair of the 12th IEEE International Conference on Computer and Information Technology and the Mobimedia 2015. He is the General Vice Chair of the Tridentcom 2014. He is a Keynote Speaker of CyberC 2012, Mobiquitous 2012, and Cloudcomp 2015.
\end{IEEEbiography}

\end{document}